\documentclass[twocolumn,showpacs, preprintnumbers,  ]{revtex4}
\usepackage{amssymb}

\usepackage{graphicx}
\usepackage{dcolumn}
\usepackage{bm}

\begin{document}

\preprint{APS/123-QED}
\title{Interplay between the magnetic fluctuations and superconductivity \\
in the lanthanum cuprates}
\author{G.\ B.\ Teitel'baum,\thanks{
E-mail: grteit@dionis.kfti.knc.ru} V.\ E.\ Kataev, E.\ L.\ Vavilova,}
%\affilation
\address{E.K.Zavoiskii Institute for Technical Physics of the RAS,\\
Sibirskii Trakt 10/7,\\
Kazan 420029, RUSSIA}
\author{P.\ L.\ Kuhns, A.\ P.\ Reyes, and W.\ G.\ Moulton}
%\affilation
\address{NHMFL, 1800 E P.Dirac Dr., \\
Tallahassee FL 32310, USA}

\date{\today }

\begin{abstract}
We report the analysis of the magnetic fluctuations in the superconducing $%
\mathrm{La_{2-x}Sr_xCuO_4}$ and the related lanthanum cuprates having the
different symmetry of the low temperature structure. The NMR and ESR
investigations revealed the dynamical coexistence of the superconductivity
and the antiferromagnetic correlations in the large part of
superconductivity region of the phase diagram. We show that for all
compounds, independent on their low temperature symmetry and on their
superconducting properties, the enhancement of the spin stiffness near 1/8
doping takes place.
\end{abstract}

\pacs{ 74.25.Ha; 74.72.Dn; 76.30.-v; 76.60.-k}
\maketitle

%\affilation
\address{E.K.Zavoiskii Institute for Technical Physics of the RAS,\\
Sibirskii Trakt 10/7,\\
Kazan 420029, RUSSIA}

%\affilation
\address{NHMFL, 1800 E P.Dirac Dr., \\
Tallahassee FL 32310, USA}

%\affilation
\address{E.K.Zavoiskii Institute for Technical Physics of the RAS,\\
Sibirskii Trakt 10/7,\\
Kazan 420029, RUSSIA}

%\affilation
\address{NHMFL, 1800 E P.Dirac Dr., \\
Tallahassee FL 32310, USA}

%   Tel.  (007/8432) 721154 ;
%   Fax.  (007/8432) 725075 ;

The interest to the microscopic phase separation in the high-$T_{c}$
superconducting  materials has received a strong impetus after the discovery
of stripe correlations \cite{gb1}. They were observed only in the compounds
specially doped with the rare earth ions whose role is to induce the low
temperature tetragonal\textrm{\ (LTT)} phase favorable for the pinning of
the stripe fluctuations. Recent neutron scattering experiments \cite{gb2} in
the low temperature orthorhombic \textrm{(LTO)} phase of $\mathrm{%
La_{2-x}Sr_{x}CuO_{4}}$ with $x=0.12$ reveal the presence of modulated
antiferromagnetic order very similar to that found in \textrm{LTT} compound $%
\mathrm{{La_{1.6-x}Nd_{0.4}Sr_{x}CuO_{4}}}$. But on the larger time scale
the magnetic fluctuations in $\mathrm{La_{2-x}Sr_{x}CuO_{4}}$ are dynamical
especially for the superconducting state and their relevance to the stripe
structure is a matter of debate. In particular, the dynamical character of
the microscopic phase separation hinders the investigation of its properties
by means of low frequency local methods such as conventional NMR \cite%
{gb3,gb4}.

The main aim of the present work is to analyze the phase diagram and the
properties of magnetic fluctuations for superconducting $\mathrm{%
La_{2-x}Sr_{x}CuO_{4}}$ and related compounds with a help of experiments
whose characteristic frequency is shifted to larger values in comparison
with the conventional NMR. We consider \textrm{ESR} ($\nu \backsim 10$ $%
\mathrm{GHz}$) and\ high field \textrm{NMR} ($\nu \backsim 0.1$ $\mathrm{GHz}
$) measurements which are focused on a comparative analysis of the magnetic
fluctuations for the different metalloxides. With this purpose we discuss
the ESR data obtained for such compounds as $\mathrm{La_{2-x}Sr_{x}CuO_{4}}$
\textrm{(LSCO)} \cite{gb5}, \ $\mathrm{La_{2-x}Ba_{x}CuO_{4}}$ \textrm{%
(LBCO) }\cite{ad3}, $\mathrm{La_{2-x-y}Eu_{y}Sr_{x}CuO_{4}}$ \textrm{(LESCO)
}\cite{ad4} together with the conventional NMR data for $\mathrm{%
La_{2-x-y}Nd_{y}Sr_{x}CuO_{4}}$ \textrm{(LNSCO) }\cite{gb8} and new high
field NMR data for superconducting \textrm{LSCO. }All the measurements were
carried out on powder samples with various hole doping. For \textrm{LSCO }%
the doping level covers the entire superconducting region of the phase
diagram, for\textrm{\ } \textrm{LBCO }we studied the doping region in the
vicinity of the well known $T_{c\text{ }}$dip, whereas the \textrm{LESCO}
and \textrm{LNSCO}\ series correspond to the nonsuperconducting \textrm{LTT }%
phase. The samples which were used for the \textrm{ESR} measurements were
doped with 1 at. \% of \textrm{Gd}, used as \textrm{ESR} probe \cite{gb5}.
Such tiny concentration of Gd ensured only the small suppression of $T_{c}$
via pair breaking.

\begin{figure}[h]
\includegraphics[width= 7 cm]{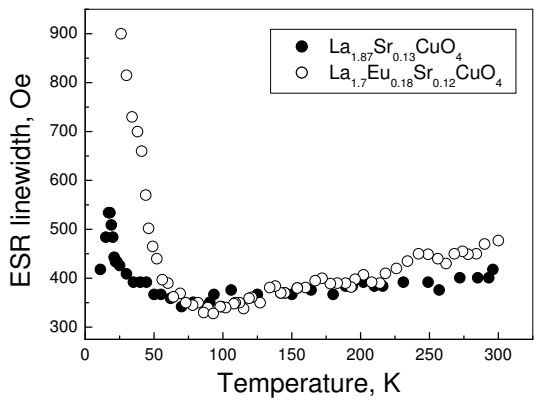}
\caption{The typical ESR linewidth temperature dependences for LTO
superconducting LSCO and LTT nonsuperconducting LESCO. }
\label{Fig1}
\end{figure}

We analyzed the temperature and concentration dependence of the width of the
most intense component of multiline $\mathrm{Gd^{3+}}$ \textrm{ESR}
spectrum, corresponding to the fine splitting of the spin states $S=7/2$ in
the crystalline electric field \cite{gb5}. The typical temperature
dependence of the linewidth $\delta H$ is shown in Fig.\ref{Fig1}.

The temperature behaviour for $T>T_{c}$ is qualitatively very similar for
all samples under study: a linear dependence of $\delta H$ \ on temperature
which is followed by the rapid growth of the linewidth at low $T$. \ But
after cooling below 40K the behaviour of superconducting and
nonsuperconducting samples becomes different: the linewidth of
superconducting \ \textrm{LSCO} exhibits the downturn starting at a
temperature $T_{m}$ dependent on $x$ whereas for other samples which are not
bulk superconductors the linewidth continues to grow upon further lowering
temperature (See Fig.1).

This behaviour may be explained if to take into account that in addition to
the important but temperature independent residual inhomogeneous broadening
the linewidth is given by different homogeneous contributions linked to the
magnetic properties of $\mathrm{CuO_{2}}$ planes:

i) the interaction of $\mathrm{Gd^{3+}}$ spins with the charge carriers,
i.e. the Korringa relaxation channel. The simplest Korringa term in the
linewidth is $\delta H=a+bT$ with $b=4\pi {(}JN_{F}{)}^{2}P_{M}$ (Ref.%
\onlinecite{gb6}), where $P_{M}=[S(S+1)-M(M+1)]$ - is the squared matrix
element of the \textrm{Gd }spin  transitions between the $M$ and $M+1$
states, $\ N_{F}$ is the density of states at the Fermi level, $\ J$ is the
coupling constant between the \textrm{Gd }and charge carriers spins \cite%
{gb5}. The factor $P_{M}$ describes the Barnes-Plefka enhancement \cite{gb6}
of the relaxation with respect to the standard Korringa rate. Such an
enhancement occurs in exchange-coupled crystal field split systems where the
g-factors of localized and itinerant electrons are approximately equal but
the relaxation of conduction electrons towards the ''lattice'' is strong
enough to inhibit bottleneck effects. For the system under study it was
discussed in Ref.\onlinecite{gb5}. Note, that the enhancement of the linear
slope for \mbox{LESCO} compound relative to that for LSCO seen in Fig.1 is due to
the influence of the depopulation of the first excited magnetic \textrm{Eu }%
level \cite{ad4}.

ii) the interaction of Gd with copper spins, giving rise to homogeneous
broadening of \textrm{Gd ESR} line (a close analogue of nuclear spin-lattice
relaxation):

\begin{equation}
\delta H=\frac{1}{2}{(\gamma H)}^{2}P_{M}\left[ \left( \tau /3\right)
+\left( 2\tau /3\right) /(1+{(\omega \tau )}^{2})\right]
\label{eq}
\end{equation}%
where $\tau $ is the magnetic fluctuations life-time, $H$ is the internal
magnetic field at \textrm{Gd} site. Following Ref.\onlinecite{gb8} we assume
the activation law for the fluctuation lifetime temperature dependence $\tau
=\tau _{\infty }\exp (E_{a}/kT)$ with $\tau _{\infty }$ being the lifetime
at the infinite temperature and $E_{a}$ - the activation energy,
proportional to the spin stiffness $\rho _{s}$ $\left( E_{a}=2\pi \rho
_{s}\right) $.

The second contribution describes the standard Bloembergen-Purcell-Pound
(BPP) behaviour: the broadening of the ESR line upon cooling  with the
downturn at certain freezing temperature $T_{m}$ corresponding to $\omega
\tau =1$. Here $\omega $ is the resonant frequency. This expression is
written for the case when the fluctuating magnetic fields responsible for%
\textrm{\ Gd} spin relaxation are induced by local \textrm{Cu} moments. In
the polycrystalline samples the averaging over the random orientation of the
local \textrm{Cu} moments with respect to the external magnetic field yields
by a factor of 2 larger probability of their perpendicular orientation as
compared to the collinear one.

\begin{figure}[h]
\includegraphics[width= 7 cm]{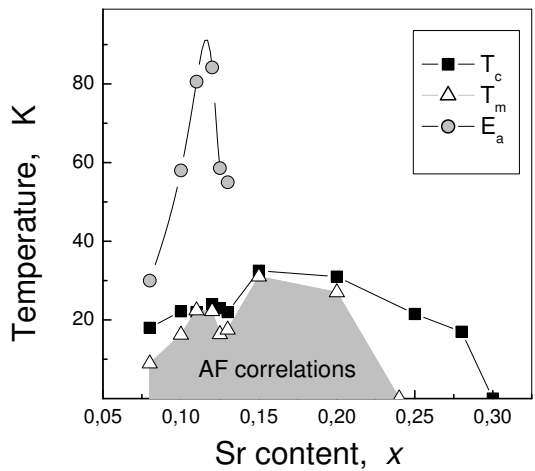}
\caption{The phase diagram of the magnetic fluctuations of superconducting
LSCO. The triangles correspond to the magnetic transition temperature $%
T_{m}(x)$, squares - to the superconducting transition temperature $T_{c}(x)$
and the circles to the magnetic fluctuations activation energy $E_{a}(x)$.
The coexistence region is shown with a grey color.}
\label{Fig2}
\end{figure}

We observed that depending on the Sr content the linewidth behaviour
transforms from the \textrm{BPP}-like (with the maximum at $T_{m}$) to the
pure Korringa (linear) temperature dependence. Basing on the observation
that the relative weight of the \textrm{BPP}-contribution, compared with the
Korringa one, decreases with increasing \textrm{Sr} doping we conclude that
at low $x$ the \textrm{Gd} spin probes almost magnetically correlated state
and at the high $x$ end - almost nonmagnetic metal. Such a transformation
may be explained it terms of the microscopical phase separation to the
metallic and AF correlated phases. It is worth to remind that very soon
after the discovery of the high $T_{c}$ superconductivity in cuprates it was
suggested \cite{gb10}, that the microscopical phase coexistence is the
inherent feature of these materials. Note that according to the phase
diagram shown in Fig.\ref{Fig2} the obtained $T_{m}$ values are lower than
the respective $T_{c}$, although for certain hole doping they are lying
close to each other. The relative amount of the AF phase falls abruptly in
the vicinity of $x=0.20$ \ so that for $x=0.24$ any traces of it are absent.
One cannot exclude that this boundary is connected with the existence of the
widely discussed quantum critical point \cite{ad7} \ at this doping \ values.

The different temperature dependences of the linewidths for the
superconducting and nonsuperconducting compounds may be consistently
explained assuming that for the superconducting samples the linewidth below $%
T_{c}$ is governed by fluctuating fields which are transversal to the
constant field responsible for the Zeeman splitting of the \textrm{Gd} spin
states (the second term in Eq.(\ref{eq}) for $\delta H$). Since these
fluctuations are induced by \textrm{Cu} moments lying in the \textrm{CuO}$_{%
\text{2}}$ planes, it means that \textrm{Gd} ions are subjected to
the constant magnetic field normal to these planes. This may
indicate that the magnetic flux lines penetrating in the layered
superconduting sample tend to orient normally to the basal planes
where the circulating superconducting currents flow (it is also
possible, that \textrm{Gd} ions pin the magnetic fluctuations
connected with the normal vortex cores). The important argument in
favor of the magnetic fluctuations contribution to the Gd ESR
linewidth is given by the fact that the BPP peak at
$T=T_{m}(x=0.10)\approx 16$ K is in a reasonable agreement with
that observed near 4 K in the $^{139}$La nuclear \ spin relaxation
rate temperature dependence for LSCO compound with $x=0.10$ at the
frequency of 140 MHz \cite{gb11}.

In principle there might be also a second possibility of the different low
temperature behaviour of the linewidth for superconducting samples in
comparison with that for nonsuperconducting ones. The nonresonant field
dependent microwave absorption in the superconducting state may distort the
shape of the \textrm{ESR} spectrum. But these distortions should be
especially pronounced for the broad lines, typically for the samples with
the small amount of holes, whereas the temperature $T_{m}$, characteristic
for small $x$, is considerably lower than $T_{c}$. Thus the possible
distortion of \textrm{ESR} lineshape owing to the nonresonant microwave
absorption as the main reason for the apparent narrowing of the \textrm{ESR}
line below $T_{c}$ seems to be improbable.

Since the measurements were carried out at nonzero external field it is very
important to consider the flux lattice effects. At typical \textrm{ESR}
fields of approximately $0.3$ \textrm{T}, oriented normally to the \textrm{%
CuO}$_{\text{2}}$ layers, the period of lattice is 860 \textrm{\AA },
whereas the vortex cores sizes for \textrm{LSCO} are approximately 20
\textrm{\AA }. As the upper critical field amounts to 62 \textrm{T}, it is
clear that in the case of ESR the vortex cores occupy only 0.5\% of the
\textrm{CuO}$_{\text{2}}$ planes. According to Ref.\onlinecite{ad5,ad6} the
\textrm{Cu} spins in the vortex cores may be \textrm{AF} ordered. Therefore
the phase diagram in Fig.2 indicates that not only the spins in the normal
vortex cores are \textrm{AF} correlated, but the \textrm{AF} correlations
are spread over the distances of the order of magnetic correlation length
which at low doping reaches 600-700 \textrm{\AA } \cite{ad6}.

Numerical simulations of the \textrm{Gd ESR} linewidths for the compounds
with the different Sr content enable us to estimate the values of the
parameters in the expression for the linewidth. For example the maximal
effective internal field $H$ in the rare earth positions is about
200\thinspace\ \textrm{Oe}; the life time $\tau _{\infty }$, which was found
to be material dependent, for LSCO is equal to $\tau _{\infty }=0.3\cdot
10^{-12}$ sec, and the activation energies $E_{a}$ for all investigated
compounds are shown in Figs.2, 3. Note that since the influence of the Nd
magnetic moments for the LNSCO compound hinders the ESR measurements the
activation energy for this compound was estimated from the measurements of
the nuclear spin relaxation on Cu and La nuclei.

\begin{figure}[h]
\includegraphics[width= 7 cm]{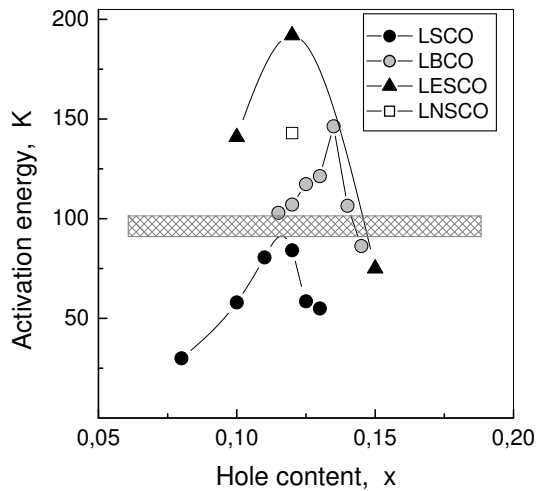}
\caption{The activation energies $E_{a}(x)\ \ $for the magnetic fluctuations
in the different cuprates versus the hole doping $x$. \ Shadowed is the
boundary separating the \ $E_{a}(x)$ \ values corresponding to
nonsuperconducting and bulk superconducting phases. }
\label{Fig3}
\end{figure}

The enhancement of $E_{a}$ (that is of a spin stiffness $\rho _{s}$) near $%
x=0.12$ shown in Fig.3 gives evidence of the developed antiferromagnetic
correlations for all investigated compounds and explains both the
anomalously narrow peak in inelastic neutron scattering \cite{gb7} and the
elastic incommensurate peak with a narrow q-width \cite{gb2} reported for
the superconducting $\mathrm{La_{2-x}Sr_{x}CuO_{4}}$ for this Sr doping.
This indicates the important role of the commensurability and gives evidence
of the plane character of the inhomogeneous spin and charge distributions.
The maximal activation energies are 80 K for \textrm{LSCO}, 144 \textrm{K}
for \textrm{LBCO}, 160 \textrm{K} for \textrm{LESCO} and 143 \textrm{K} for
\textrm{LNSCO}. Note that for \textrm{LBCO} and \textrm{LESCO} the
signatures of the bulk superconductivity \cite{ad3,ad4} become visible upon
the suppression (in course of the \textrm{Ba }or \textrm{Sr} doping) of the
activation energy down to 80-85 \textrm{K}. Therefore it is plausible to
assume that these values of the activation energy are probably the critical
ones for the realization of the bulk superconducting state. The
corresponding boundary is shown in Fig.\ref{Fig3}. Fluctuations with the
higher activation energies (spin stiffness) are effectively pinned and
suppress the superconductivity.

\begin{figure}[h]
\includegraphics[width= 7 cm]{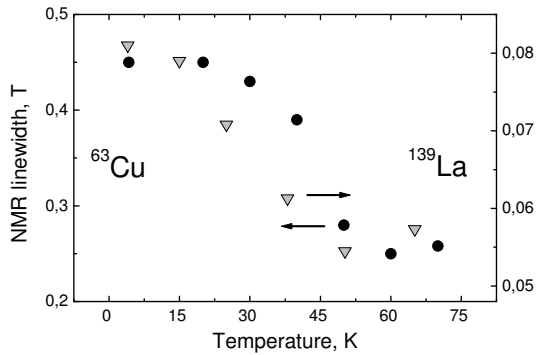}
\caption{The temperature dependence of the $^{63}$Cu and $^{139}$La NMR
linewidths for the superconducting LSCO with $x=0.12$. }
\label{Fig4}
\end{figure}

To obtain the information about the ordered magnetic moments for the \
compounds with the enhanced spin stiffness the NMR measurements at
20-25\thinspace\ T were carried out in a high homogeneity resistive magnet
of the NHMFL in Tallahassee FL. The temperature and doping dependencies of $%
^{63,65}$Cu and $^{139}$La NMR field sweep spectra of the oriented powders $%
\mathrm{La_{2-x}Sr_{x}CuO_{4}}$ were studied. According to the previous La
NQR results \cite{gb3,gb11} the measurements of oriented powder samples in a
magnetic field perpendicular to \textbf{c} axis revealed that for Sr content
near 1/8 the central lines of the observed spectra both for Cu and La
exhibit the broadening upon cooling below 40-50\thinspace\ K (Fig.\ref{Fig4}%
).

Such a behaviour is connected with the slowing down of the magnetic
fluctuations, which are gradually slowing down upon ordering. The broadening
of the La NMR line allows us to estimate that the additional magnetic field
at La nucleus is 0.015\thinspace\ T. If we consider that for the
antiferromagnet $\mathrm{La_{2}CuO_{4}}$ the copper moment of 0.64$\mu _{B}$
induces at the La site the field of 0.1\thinspace \textrm{T} \cite{gb12},
then the effective magnetic moment in the present case is $\sim 0.09\mu _{B}$%
. Note that the manifestation of the magnetic order only in the vicinity of $%
x=1/8$,  when the AF structure is commensurate with the lattice, indicates
that the magnetic inhomogeneities are of a plane character.

In conclusion our investigation reveals that for all studied compounds
independent on the symmetry type (LTO or LTT) in the neighbourhood of $1/8$
doping the enhancement of the spin stiffness takes place. The compounds with
the spin stiffness larger than the certain critical value (See Fig.3) reveal
no bulk superconductivity.

According to the phase diagram the inherent feature of the superconducting
state in cuprates is the presence of frozen antiferromagnetic correlations.
Such a coexistence seems to be a result of phase separation at the
microscopic scale as it was discussed in pioneering paper of Gor'kov and
Sokol \cite{gb10}.

In the neighbourhood of $1/8$ doping this coexistence may be realized in a
form of dynamic stripes, since the corresponding enhancement of the
spin-stiffness reveals the plane character of the spin (and charge)
inhomogeneities.

One of the authors (G.B.T.) is grateful to L.P.Gor'kov for the valuable
discussions of the phase separation specifics for the cuprates. This work is
partially supported through the RFFR Grant N 01-02-17533.

\end{document}